\documentclass{aa} 
\usepackage{graphicx}
\usepackage{natbib}
\bibpunct{(}{)}{;}{a}{}{,} 

\begin{document}
  \title{Interpreting  the  spectral behavior of  MWC\,314
    \thanks {Based on observations made at the 0.91\,m of Catania Observatory, the OHP telescopes and the 1.83\,m telescope of the Asiago Observatory.}    }
   \author{A. Frasca\inst{1}\and
    A. S. Miroshnichenko\inst{2}\and
    C. Rossi\inst{3}\and
    M. Friedjung\inst{4}\thanks{Michael Friedjung passed away on October 22, 2011.}\and
    E. Marilli\inst{1}\and
    G. Muratorio\inst{5}\and
    I. Bus\`a\inst{1}
             }

\offprints{A. Frasca\\ \email{antonio.frasca@oact.inaf.it}}

\institute{INAF, Osservatorio Astrofisico di Catania, via S. Sofia, 78, 95123 Catania, Italy
    \and
    Department of Physics and Astronomy, University of North Carolina at Greensboro, Greensboro, NC
    27412, U.S.A
    \and
Department of Physics, University La Sapienza, Piazzale A. Moro 5, 00185  Roma, Italy
    \and
    Institut d'Astrophysique CNRS+Paris 6, Paris, France
    \and
    OMP/LAM, Marseille, France
             }

   \date{Received   ; accepted }

  \abstract
   {  MWC\,314 is one of the most luminous stars in the Milky Way.
   Its fundamental parameters are similar to those of LBVs, although no large photometric variations have been recorded.
   Moreover, it shows no evidence neither for a dust shell nor for relevant spectral variability.
    }
   { The main purpose of this work is to clarify the origin of the radial velocity and line profile variations exhibited by
     absorption and emission lines.
     }
   { We  analyzed  the radial velocity (RV) variations displayed by the absorption lines from the star's atmosphere using high resolution  optical spectra and fitting the RV curve with an eccentric orbit model.
  We also studied the RV and profile variations of some  permitted and forbidden emission lines of metallic ions with a simple geometric model.
The behavior of the Balmer and \ion{He}{i} lines has been investigated as well.
    }
   { Fourier analysis applied to the RV of the absorption lines clearly shows a 60--day periodicity.  A dense coverage
   of the RV curve allowed us to derive accurate orbital parameters.  The RV of the  \ion{Fe}{ii}
   emission lines varies in the same way, but with a smaller amplitude. Additionally, the  intensity ratio of the blue/red peaks of these
   emission lines correlates with the RV variations. The first three members of the Balmer series as well as [\ion{N}{ii}] lines
   display a nearly constant RV and no profile variations in phase with the orbital motion instead.
   The \ion{He}{i}\,$\lambda$5876 \AA\ line shows a strongly variable profile with broad and blue-shifted absorption components
   reaching velocities of $\le -1000$\,km\,s$^{-1}$ at some specific orbital phases.
   }
 { Our data and analysis provide strong evidence that the object is a binary system composed from a supergiant B[e] star
and an undetected companion.
The emission lines with a non-variable RV could originate in a circumbinary region.
 For the \ion{Fe}{ii} emission lines  we propose a simple geometrical two-component model where a compact source of \ion{Fe}{ii} emission, moving around the center of mass, is affected by a static extra absorption that originates from a larger area. Finally, the blue-shifted absorption in the \ion{He}{i}\,$\lambda$5876 \AA\ line could be the result of density enhancements in the primary
 star wind flowing towards the companion that is best observed when projected over the disk of the
primary star.
 }

   \titlerunning{The spectral behavior of MWC\,314}
    \authorrunning{A. Frasca et al.}

   \maketitle
\section{Introduction}
\label{Sec:intro}
\begin{figure*}[htp]
\hspace{-1.cm}
\includegraphics[width=18cm]{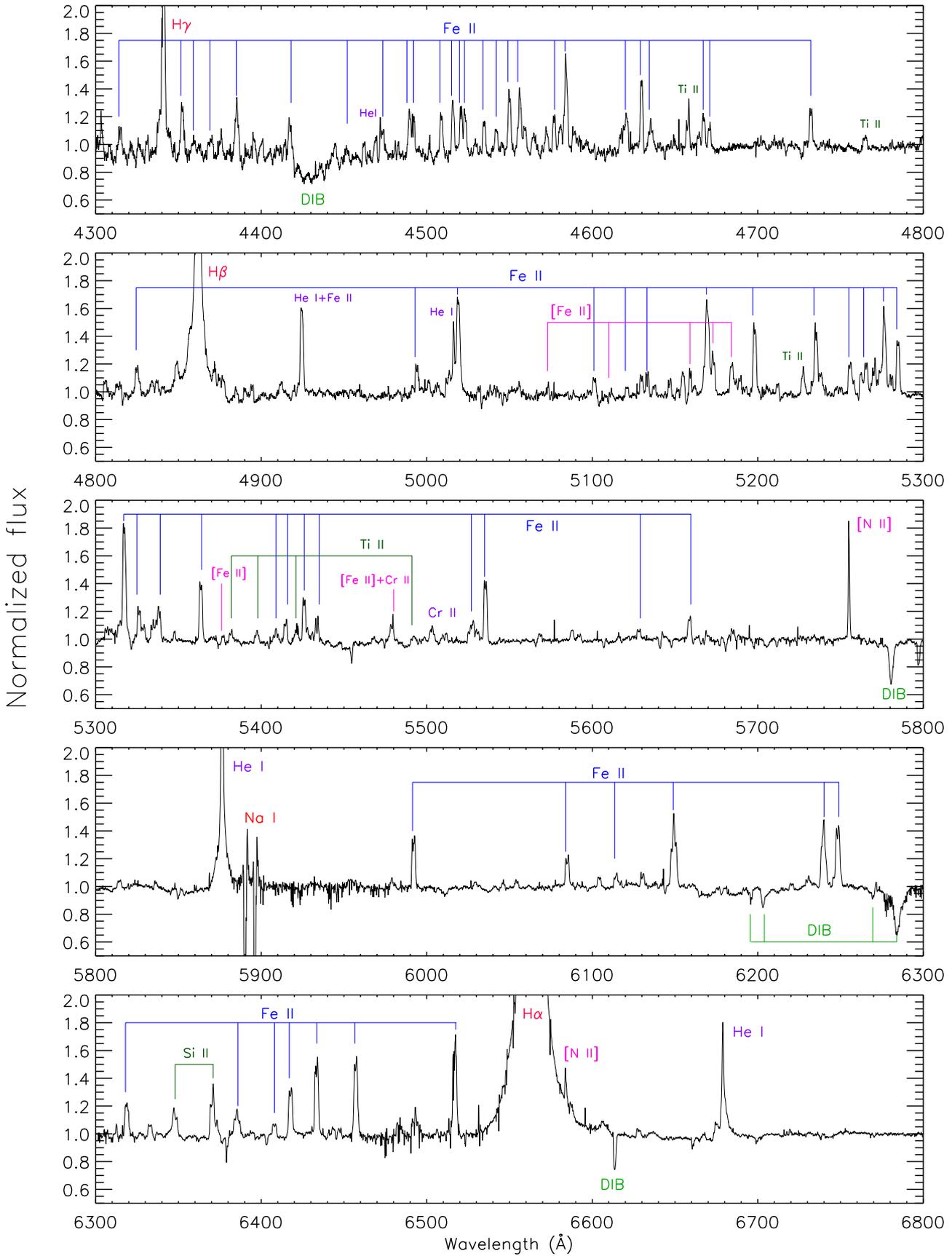}
\caption{Continuum-normalized {\sc Elodie}
spectrum taken in August 1995 with the most prominent emission and
absorption lines identified. All the absorption lines visible
between the \ion{Na}{I}\,D doublet and 6000\,\AA\ and those on the
blue side of the H$\alpha$ line have telluric origin.
 } 
\label{fig:Elodie_spe}
\end{figure*}

There has been debate about whether \object{MWC\,314} is a B[e] supergiant or
a luminous blue variable (LBV). It appears to be one of the most
luminous stars in the Galaxy \citep{Miro98}. However, it does not
show the typical brightness variations of the latter class. The
profiles of the permitted emission lines show two peaks, suggesting
their formation in a rotating disk which led us to consider MWC\,314
a B[e] supergiant rather than an LBV \citep{Murat08}.

\citet{Marston08} detected a very large bipolar
nebula that emits in H$\alpha$ and extends for $\sim$13.5\,pc. It
takes over $10^5$ years to form such a nebula at a typical expansion
velocity for LBV candidates of 50\,km\,s$^{-1}$ \citep{Nota95}. They
propose that this star and other B[e] supergiants with large nebulae
could have passed the LBV phase and have been moving towards the
blue part of the upper HR diagram.
This possibility is also supported by the absence of circumstellar dust,
analogously to \object{P~Cyg}, already noted by \citet{Miro96} and
confirmed by the spectral energy distributions of these two
stars shown by \citet{Lobel13}.  This might mean that the dust has
been blown away by the wind and one or more LBV outbursts, which
occurred long ago. 

The possibility of binarity of MWC\,314 was suggested by
\citet{Wisn06} from fast variations of the H$\alpha$ emission line
strength. It was further indicated by the discovery of regular
displacements of the absorption line centroids in the spectrum of
MWC\,314 \citep{Murat08}.

Regular radial velocity (RV) variations of four absorption lines with a
period of 60.8 days were detected by \citet{Lobel13} based on 15
high-resolution optical spectra. These authors concluded that
MWC\,314 is a single-lined binary system with a LBV-like early
B--type supergiant primary with a mass of $\sim$40 M$_\odot$ that
has a strong asymmetric wind. No signature of the secondary
companion was found, although they suggested that it might be a less
massive relatively cool (T$_{\rm eff} \sim$ 6200 K) giant.

\citet{Richardson14} presented a larger set of
high-resolution spectroscopic data that cover the period 2001--2013
and derived a slightly different orbit using the same absorption
lines as \citet{Lobel13}. However, they concluded that the primary
star in the system has a mass of $\sim$5 M$_{\odot}$. Such a low
mass was explained by a strong mass loss through transfer toward the
secondary, stellar wind, and eruptions that led to the mass ratio
reversal. At the same time, these authors did not reconsider the
primary's luminosity, thus suggesting that it is significantly
overluminous for its current mass.

They also obtained $K'$-band interferometry that was modelled with a point-like source 
plus an elliptical Gaussian corresponding to a circumbinary disk tilted by 50--65$\degr$ 
with respect to the line of sight. However, they do not exclude that this structure is a jet 
roughly aligned with the H$\alpha$ bipolar outflow. 

In any case, the lack of detection of spectral
signatures from the secondary component does not permit establishing
more accurate masses of the stellar components.

In order to study this binary system with a large set of data and to look for correlations
of permitted and forbidden emission lines with the orbital phase, we performed a dense
spectroscopic monitoring of MWC\,314 between September 2007 and October 2009.

In this paper we  describe the results obtained from the analysis of
all the spectra collected from 1994 to 2009. We studied the  RV behavior
of both absorption and emission lines, the simultaneous ratio of blue/red peak intensities
of permitted emission lines, and the total equivalent width of the emission lines.

\section{Observations and data reduction} \label{Sec:Data}

The first set of spectra was obtained between 1994 and 1998 at OHP  with both the
{\sc Aurelie} and {\sc Elodie} spectrographs and one spectrum at Ekar-Asiago in 2006  with
the {\sc Reosc} \'echelle spectrograph.  A detailed logbook  of
these observations and the data reduction procedures are described
in \citet{Murat08}.

Additional spectra were collected between September 2007 and  October 2009 in
the wavelength range 4300--6850\,\AA\  at the 91\,cm telescope of  the
Catania Observatory (OAC) with the {\sc Fresco} spectrograph giving a
spectral resolving power of  $R\simeq21$\,000.
In some cases we were able to follow the star for several
consecutive nights.

The data were reduced with the {\sc echelle} task of the IRAF
package following standard steps. The data analysis was performed
with IRAF routines and IDL\footnote{IDL (Interactive Data Language) is a registered
trademark of Exelis Visual Information Solutions.} procedures.

The log of all the spectroscopic observations is presented in
Table~\ref{tab:rv}, where also the numerical results of our analysis
are reported.

An {\sc Elodie}  spectrum showing the most prominent emission lines in the range 
4300--6850\,\AA\ is shown in Fig.\,\ref{fig:Elodie_spe}.

\section{Data analysis}
\label{Sec:analysis} We studied the behavior of absorption and
emission (permitted and forbidden) lines from several points of
view.

 We used the absorption lines only to measure the RV with the aim to 
confirm the strong variations detected in a relatively short time
\citep{Murat08} and to refine the parameters of the periodicity
previously found on the basis of a small number of observations
spanning a period of several years. In the hypothesis of a binary
system, possibly composed from two very luminous stars, we studied
the emission lines aiming at understanding not only their forming
regions and the cause of their double-peaked shape, but also at
interpreting their temporal variations and looking for any possible
correlation with the RV of the absorption lines.

  \subsection{Absorption lines}
\begin{figure}[htp]
\hspace{-1.cm}
\includegraphics[width=10.2cm]{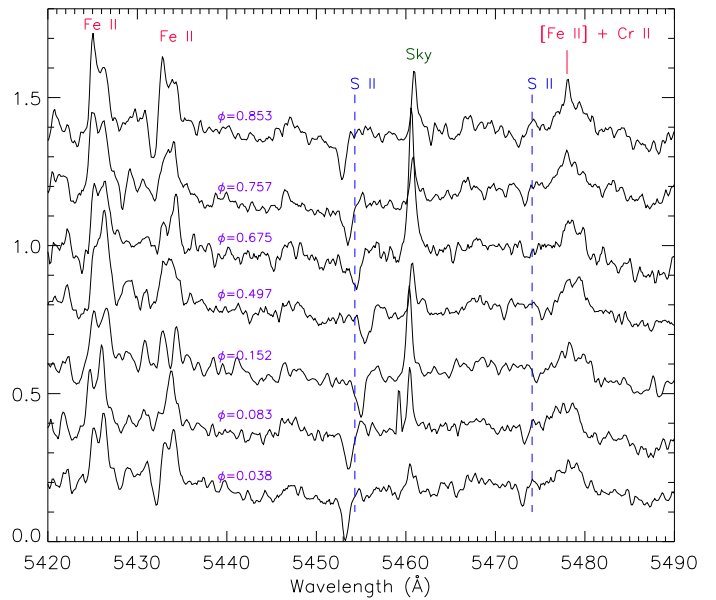}
\caption{Example of seven FRESCO spectra in the range around 5450\,\AA.
The velocity shift of the absorption lines (\ion{S}{ii}\,$\lambda \lambda 5453.8$ and $\lambda 5473.6$\,\AA) is apparent.
Also note the variable  intensity ratio of the blue and red peaks of the emission line \ion{Fe}{ii} $\lambda 5425.3$\,\AA.
 }
\label{fig:abs_lines}
\end{figure}

All the absorption lines are very faint, therefore  we carefully
selected a number of them to be used for the RV determination. We
chose lines of the same ion and possibly same multiplet previously
noticed for their wavelength clearly varying in time \citep[see,
e.g.,][and Fig.~\ref{fig:abs_lines} of the present paper]{Rossi11}.
  The final list of these lines is as follows:
  \ion{S}{ii} ($\lambda 5201.0$, $\lambda 5212.6$, $\lambda 5320.7$, $\lambda 5453.8$,
  $\lambda 5473.6$,  $\lambda 5556.0$, $\lambda 5564.9$, $\lambda 5606.1$, $\lambda 5640.0$,
  $\lambda 5647.0$),
\ion{N}{ii} ($\lambda5676.0$, $\lambda5679.6$, and $\lambda5710.8$),
and \ion{Ne}{i} ($\lambda6402.2$). A few other lines were also
considered in the spectra with the highest S/N ratio.

For each line we have measured the centroid wavelength and converted
it into the heliocentric radial velocity. For each spectrum we then
computed the average and the standard deviation of all the
individual RV values. These values are reported in
Table~\ref{tab:rv}.
\begin{table*}
\caption{Radial velocity of absorption lines ($RV_{\rm abs}$),
blue/red peak intensity ratio ($B/R$) velocity of the blue and red
emission peaks and radial velocity of single emissions (see text)}
\label{tab:rv}
\begin{tabular}{l l r r c c c r r c c c c}
\hline\hline 
\noalign{\smallskip}
 Date  & HJD & $RV_{\rm abs}$ & err & phase & $B/R$ & err &  $RV_{\rm em}^{\rm B}$ & err & $RV_{\rm em}^{\rm R}$ & err & $RV_{\rm em}^{\rm S}$  & err \\
  & 2\,400\,000+ &  \multicolumn{2}{c}{(km\,s$^{-1}$)} & & & & \multicolumn{2}{c}{(km\,s$^{-1}$)} &  \multicolumn{2}{c}{(km\,s$^{-1}$)}   &  \multicolumn{2}{c}{(km\,s$^{-1}$)}  \\
\noalign{\smallskip}
\hline
\noalign{\smallskip}
1994 08 28 &  49593.39754  &  $-25$ &6  &  0.780  &   1.15  & 0.12  &  $ -6.7$   & 2.4  &   59.6  &  2.6 & 27.2 & 2.7 \\
1995 06 07 &  49876.58716  &  $ 97$  &4  &  0.443  &   0.55  & 0.11  &  $ -8.2$   & 1.8  &   55.1  &  3.1 & 42.9 & 4.3 \\
1995 08 19 &  49949.46725  &  $ 47$  &5  &  0.643  &   0.87  & 0.06  &  $ -0.6$   & 3.1  &   63.7  &  2.7 & 37.4 & 3.7 \\
1998 07 05 &  51000.46328  &  $-80$ &7  &  0.947  &   ...   &   ...  &  $  ... $   & ...  &   ...   &  ... &   ...      &  ... \\
1998 07 21 &  51016.46947  &  $ 95$  &9  &  0.210 &   ...   &   ...  &  $  ... $   & ...  &   ...   &  ... &   ...  &  ... \\
1998 09 09 &  51066.35217  &  $-38$ &5  &  0.032  &   1.06  & 0.04  &  $ -2.0$   & 5.7  &   59.8  &  3.1 &34.9  & 3.5 \\
2006 07 16 &  53933.40647  &  $ 85$  &6  &  0.236  &   0.76  & 0.03  &  $ -5.7$   & 4.3  &   68.4  &  3.4 &40.9  & 4.1  \\
2007 09 03 &  54347.40815  &  $ ... $  & ...  &  0.052  &   0.77  & 0.08  &  $ 13.1$   & 2.0  &   81.0  &  3.0 &36.4   & 3.6 \\
2007 09 29 &  54373.38428  &  $ 85$  &8  &  0.480  &   0.55  & 0.10  &  $ -5.8$   & 1.1  &   58.4  &  0.7 &42.5  & 4.2 \\
2007 09 30 &  54374.35878  &  $ 85$  &7  &  0.496  &   0.50  & 0.12  &  $ -8.8$   & 1.6  &   57.9  &  0.7 &43.1  & 4.3 \\
2008 05 17 &  54604.50475  &  $105$ &8  &  0.285  &   0.74  & 0.11  &  $  3.6$   & 3.6  &   73.5  &  4.5 &39.9  & 3.9 \\
2008 05 21 &  54607.54430  &  $102$ &1  &  0.335  &   0.66  & 0.07  &  $ -6.2$   & 2.8  &   65.4  &  1.5 &43.2  & 4.3 \\
2008 05 23 &  54609.56085  &  $104$ &4  &  0.368  &   0.65  & 0.11  &  $ -3.6$   & 4.7  &   66.5  &  3.9 &46.3  & 4.6 \\
2008 05 25 &  54611.56548  &  $102$ &5  &  0.401  &   0.58  & 0.07  &  $ -7.6$   & 2.5  &   61.6  &  5.5 &43.6  & 4.3  \\
2008 05 27 &  54613.53195  &  $ 94$  & 6 &  0.434  &   0.58  & 0.12  &  $ -9.5$   & 4.4  &   58.9  &  2.5 &45.3  & 4.5  \\
2008 06 13 &  54631.45182  &  $ 12$ & 8 &  0.729  &   1.00  & 0.05  &  $ -5.6$   & 3.6  &   62.4  &  3.2 &30.9    & 3.0  \\
2008 06 15 &  54632.55436  &  $  6$  &10 &  0.747  &   1.13  & 0.15  &  $-2.0 $   & 2.7  &   63.0  &  3.8 &29.3  & 2.9  \\
2008 06 19 &  54636.52365  &  $-11$  & 4 &  0.812  &   1.15  & 0.17  &  $ -7.2$   & 5.1  &   65.0  &  6.8 &22.0  & 2.2 \\
2008 07 20 &  54668.51706  &  $ 90$  &10 &  0.339  &   0.73  & 0.03  &  $ -1.0$   & 1.2  &   69.0  &  1.5 &42.5  & 4.2 \\
2008 07 21 &  54669.48507  &  $108$  &20 &  0.355  &   0.66  & 0.08  &  $ -4.9$   & 6.2  &   65.2  &  2.1 &44.2  & 4.4  \\
2008 07 22 &  54670.43017  &  $103$  &20 &  0.371  &   0.65  & 0.15  &  $ -6.7$   & 3.4  &   64.2  &  4.1 &45.8  & 4.6  \\
2008 07 23 &  54671.42815  &  $ 92$  &7  &  0.387  &   0.62  & 0.06  &  $-11.4$   & 3.1  &   62.4  &  2.0 &44.7  & 4.4  \\
2008 07 31 &  54679.46166  &  $ 77$  &7  &  0.519  &   0.53  & 0.12  &  $ -9.0$   & 3.7  &   58.7  &  3.7 &48.1  & 4.8  \\
2008 08 01 &  54680.46382  &  $ 79$  &8  &  0.536  &   0.56  & 0.10  &  $-10.0$   & 3.6  &   62.8  &  3.6 &46.3  & 4.6  \\
2008 08 30 &  54709.46967  &  $-43$  &8  &  0.013  &   1.11  & 0.07  &  $ -4.2$   & 1.1  &   60.6  &  1.5 &30.1  & 3.0   \\
2008 09 22 &  54732.34722  &  $ 95$  &7  &  0.390  &   0.60  & 0.08  &  $ -7.0$   & 4.6  &   65.2  &  3.1 &47.2  & 4.7   \\
2008 09 28 &  54738.39732  &  $ 92$  &10 &  0.490  &   0.63  & 0.08  &  $ -3.0$   & 5.0  &   59.3  &  0.9 &45.6  & 4.5  \\
2009 06 13 &  54996.45406  &  $  8$  &8  &  0.738  &   1.05  & 0.09  &  $ -2.1$   & 3.4  &   67.0  &  3.2 &34.9  & 3.5  \\
2009 06 15 &  54997.57071  &  $ -9$  &8  &  0.757  &   1.23  & 0.10  &  $ -7.2$   & 3.3  &   61.0  &  3.9 &27.2  & 2.7  \\
2009 06 16 &  54998.53984  &  $-15$  &7  &  0.773  &   1.15  & 0.14  &  $ -7.4$   & 2.9  &   58.2  &  4.0 &28.4  & 2.8 \\
2009 06 20 &  55003.39754  &  $-57$  &7  &  0.853  &   1.32  & 0.06  &  $-10.4$   & 3.6  &   55.5  &  5.0 &20.5  & 2.0   \\
2009 06 20 &  55003.43507  &  $-51$  &8  &  0.853  &   1.38  & 0.09  &  $ -9.5$   & 3.6  &   57.1  &  4.2 &21.6  & 2.1  \\
2009 06 24 &  55006.53995  &  $-68$  &4  &  0.905  &   1.31  & 0.11  &  $-12.3$   & 3.5  &   50.9  &  3.5 &19.0  & 1.9   \\
2009 06 24 &  55007.48101  &  $-77$  &3  &  0.920  &   1.27  & 0.08  &  $-12.8$   & 4.0  &   49.9  &  5.2 &21.5  & 2.1   \\
2009 06 25 &  55007.54341  &  $-68$  &5  &  0.921  &   1.40  & 0.12  &  $-13.0$   & 3.5  &   53.0  &  3.5 &20.7  & 2.1   \\
2009 06 25 &  55008.48148  &  $-76$  &7  &  0.936  &   1.29  & 0.14  &  $ -9.2$   & 3.8  &   56.7  &  3.2 &21.7  & 2.2   \\
2009 06 26 &  55009.48715  &  $-77$  &7  &  0.953  &   1.24  & 0.14  &  $-10.1$   & 1.9  &   56.5  &  2.7 &21.3  & 2.1  \\
2009 07 04 &  55017.36579  &  $  6$  &8  &  0.083  &   0.82  & 0.08  &  $-12.0$   & 3.1  &   63.6  &  3.4 &35.3  & 3.5   \\
2009 07 30 &  55042.51432  &  $ 85$  &7  &  0.497  &   0.61  & 0.07  &  $ -7.5$   & 0.5  &   56.5  &  1.0 &41.5  & 4.1   \\
2009 08 09 &  55053.35448  &  $ 23$  &8  &  0.675  &   0.85  & 0.07  &  $ -3.6$   & 3.6  &   64.3  &  2.5 &35.1  & 3.5  \\
2009 08 09 &  55053.39778  &  $ 30$  &7  &  0.676  &   0.85  & 0.07  &  $ -3.7$   & 3.8  &   64.3  &  3.6 &35.5  & 3.5   \\
2009 08 31 &  55075.41518  &  $-31$  &7  &  0.038  &   0.90  & 0.08  &  $ -7.3$   & 1.9  &   59.0  &  2.3 &30.3  & 3.0   \\
2009 09 03 &  55078.36655  &  $ 16$  &8  &  0.087  &   0.86  & 0.11  &  $-12.5$   & 2.9  &   63.2  &  2.7 & 31.3 & 3.1 \\
2009 09 07 &  55082.32460  &  $ 55$  &8  &  0.152  &   0.82  & 0.07  &  $-10.4$   & 4.8  &   68.4  &  2.9 & 34.6 & 3.4  \\
2009 09 26 &  55101.43383  &  $ 89$  &8  &  0.467  &   0.57  & 0.08  &  $ -8.4$   & 3.2  &   61.9  &  1.2 & 45.2 & 4.5  \\
2009 09 26 &  55101.48352  &  $ 93$  &8  &  0.468  &   0.58  & 0.11  &  $-10.4$   & 3.8  &   61.7  &  3.9 & 45.4 & 4.5  \\
2009 10 04 &  55109.34247  &  $ 61$  &8  &  0.597  &   0.66  & 0.12  &  $-12.6$   & 1.8  &   57.2  &  4.2 & 38.4 & 3.8  \\
2009 10 11 &  55116.39065  &  $ 16$  &8  &  0.713  &   0.91  & 0.07  &  $ -5.9$   & 3.9  &   62.6  &  4.3 & 33.1 & 3.3   \\
\hline
\end{tabular}
\end{table*}

\begin{figure}[htp]
\hspace{0.5cm}
\includegraphics[width=8.0cm ]{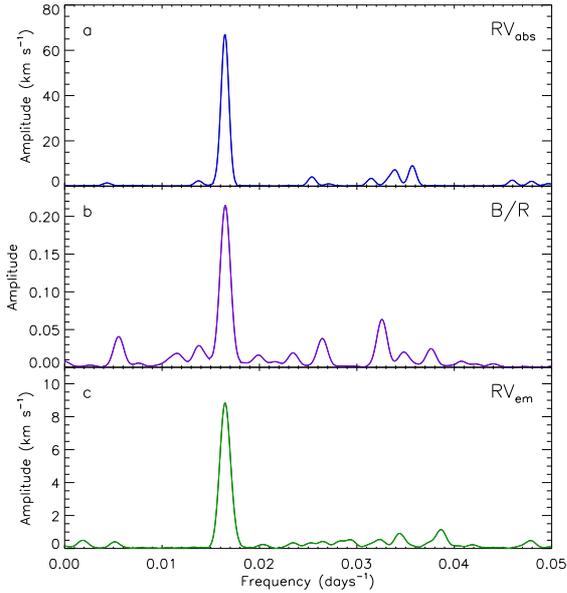}
\caption{Cleaned periodograms for $RV_{\rm abs}$ (\textit{top
panel}), $B/R$ peak ratio (\textit{middle panel}), and $RV_{\rm em}$
(\textit{lower panel}).  The 60 days period is clearly evident in
all diagrams. } \label{fig:clean}
\end{figure}

\begin{figure}[htp]
\hspace{-0.5cm}
\includegraphics[width=9.5cm]{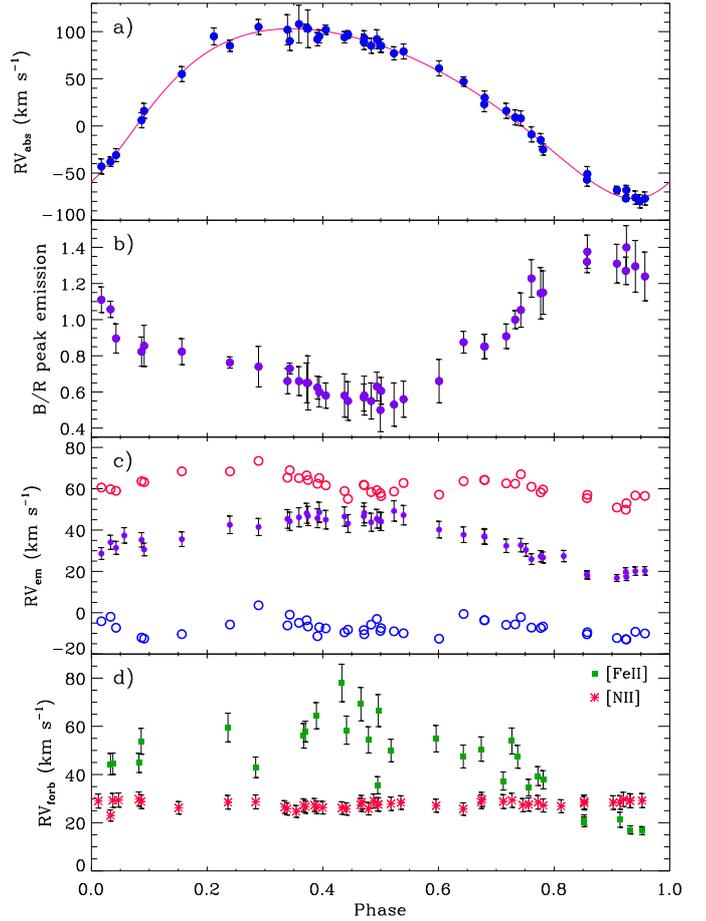}
\vspace{-0.3cm} \caption{\textit{From Top to Bottom: } a) Radial
velocity curve of MWC\,314 obtained with the absorption lines
(dots), with the RV solution overplotted as a continuous line. ~~b)
The intensity ratio of the blue to the red peak of the \ion{Fe}{ii}
emission lines.
 ~~c) Radial velocity of the \ion{Fe}{ii} emission lines as measured from their centroids (filled dots)
 and as evaluated from the red and blue
 emission peaks (upper and lower empty circles, $RV_{\rm em}^{\rm R}$  ~and~ $RV_{\rm em}^{\rm B}$).
 ~~d) Radial velocity of the forbidden [\ion{Fe}{ii}] and [\ion{N}{ii}] lines.
}
\label{fig:RV}
\end{figure}

Thanks to a large number of spectra taken in a long-term time span
(from 1994 to 2009) we were able to obtain an accurate determination
of the period of the RV variations. For this task we used
periodogram analysis \citep{sca} and the CLEAN deconvolution
algorithm \citep{rob}, which allowed us to reject aliases generated
by the spectral window of the data. We found a period of
60.8$\pm$1.7 days (see Fig.~\ref{fig:clean}). The data folded with
the period display a smooth variation with an asymmetrical shape
typical for an eccentric RV orbital motion (see
Fig.~\ref{fig:RV}-a). Thus, we fitted the observed RV curve with the
{\sc Curvefit} routine \citep{Bevi69} to determine the orbital
parameters and their standard errors.
This has also allowed us to improve the determination of the orbital period based on
the periodogram analysis.
The orbital parameters resulting from the fit of the RV curve are presented in
Table~\ref{tab:orb}, and the phase computed for each observation is
quoted in column 5 of Table~\ref{tab:rv}.

With our 48 spectra, which provide a well phase coverage of the orbital 
cycle span $\sim$15 years, we find orbital elements in a fairly good 
agreement with those derived by \citet{Lobel13}. 
Our orbital period is only 0.1\,\% smaller, the
eccentricity is $\sim$4\,\% larger, and the RV semi-amplitude is
6\,\% larger than those of \citet{Lobel13}. The RV errors of our
data are similar to those quoted by these authors, but we derive
larger errors for the orbital parameters. In particular, we find an
error of 11 minutes for the orbital period, much larger than that of
only 1.2 seconds reported by \citet{Lobel13}. We think that our
value of the orbital period error, which is not affected by possible
errors due to the relatively large photometric uncertainties, is
more reliable. 

Our orbital solution derived from the absorption lines coincides with that of
\citet{Richardson14} within the uncertainties, although our error for the orbital period is
three times smaller.
For a direct comparison of our results with those of \citet{Lobel13} and \citet{Richardson14} we also
report these latter in Table~\ref{tab:orb}. 

From our values of $P$, $k$, and $e$, we have computed the mass function, defined as:

\begin{eqnarray}
 f(m)= \frac{m_2^3}{(m_1+m_2)^2}\sin i^3 = k^3\frac{P}{2\pi G}(1-e^2)^\frac{3}{2}.
\end{eqnarray}

We find $f(m)=4.1\pm0.4$ M$_{\odot}$ (where the
uncertainty includes the errors on $P$, $k$, and $e$) in very good
agreement with the value of 4.0$\pm$0.3 M$_{\odot}$ reported by
\citet{Richardson14}. We note that without further constraints, such
as the detection of spectral features from the secondary component
and a very precise photometry, no other reliable parameter for the
system components can be obtained.

\begin{table*}
\caption{Orbital parameters derived from the RV of the absorption and emission lines. Standard errors are reported in parentheses.}
\label{tab:orb}
\begin{tabular}{l l l l l l l}
\hline\hline 
\noalign{\smallskip}
  $P_{\rm orb}$   &   HJD$_{\rm P}$  &  $e$  &  $\omega$  &   $\gamma$ &   $k$   & Diagnostics/Ref. \\
  (days)   & ($-2\,400\,000 $) &    &    &  (km\,s$^{-1}$)  &   (km\,s$^{-1}$)    \\
\noalign{\smallskip}
\hline
\noalign{\smallskip}
 60.737  (0.008)   & 49546.01  (1.10)   & 0.244  (0.020)  & 218.7$\degr$  (5.7)   & 30.7 (1.3)  & 89.7  (2.0)  &  Absorption lines   \\
60.74 (0.03)   & 49546 (2)   &  0.25 (0.05)  &  170$\degr$  (10)  &  35.0 (1.0)  &  13.5  (1.0)  &   \ion{Fe}{ii} emission lines    \\
60.799977 (0.000014) & 54959.76 (0.56) & 0.235 (0.003) & 289$\degr$ (1)  &  28.44 (0.17) &  84.5   & \citet{Lobel13}   \\
60.735 (0.024) & 55618.49 (0.65) & 0.29 (0.02) & 206.4$\degr$ (4.2)  &  31.3 (1.1) &  89.6 (1.8)  & \citet{Richardson14}  \\
\hline
\end{tabular}
\end{table*}
%
%

\subsection{Permitted emission lines}

Permitted metallic emission lines exhibit obvious variability of the blue/red peak ratio
(see Fig.~\ref{fig:abs_lines}). Taking into account the possibility
of a connection with the absorption lines velocities, we decided to
tackle the problem from several points of view. To the purpose we
selected a number of strong and isolated lines of metallic ions. The
final list includes:
  \ion{Fe}{ii}  $\lambda 6456.4$, $\lambda 6432.7$, $\lambda 6416.9$,
  $\lambda 6247.6$, $\lambda 6084.1$, $\lambda 5991.4$, $\lambda 5534.8$, and 
$\lambda 5425.3$ \,\AA~ which all display a double-peaked or asymmetric profile in all
spectra. We note that some \ion{Fe}{ii} lines were discarded due to either blending with 
other features, their faintness, or a low S/N ratio in the blue spectral range. 

First we performed a two-Gaussian fit to each line in order to
measure the intensity and wavelength of the blue and red peaks. The
results are presented in  Table~\ref{tab:rv}. We found a modulation
of the blue/red peak intensity ratio ($B/R$) with the same period as
that of $RV_{\rm abs}$. Indeed, the CLEAN analysis performed on
these ratios gives a period of 60.6$\pm$2.2 days (see
Fig.~\ref{fig:clean}) which is the same as that of $RV_{\rm abs}$
within the uncertainties. The $B/R$ curve and the $RV_{\rm abs}$ one
are nearly anti-correlated, as can be seen in Fig.~\ref{fig:RV}. The
shape of the two curves is slightly different: the minimum of $B/R$
is observed later than the maximum of $RV_{\rm abs}$, while the
largest $B/R$ ratios are observed just before the minimum of
$RV_{\rm abs}$. The high degree of correlation of the two curves is
demonstrated by a very high rank-correlation \citep{Press86}
coefficient $\rho= -0.82$.

\begin{figure}[htp]
\hspace{-0.3cm}
\includegraphics[width=9.cm]{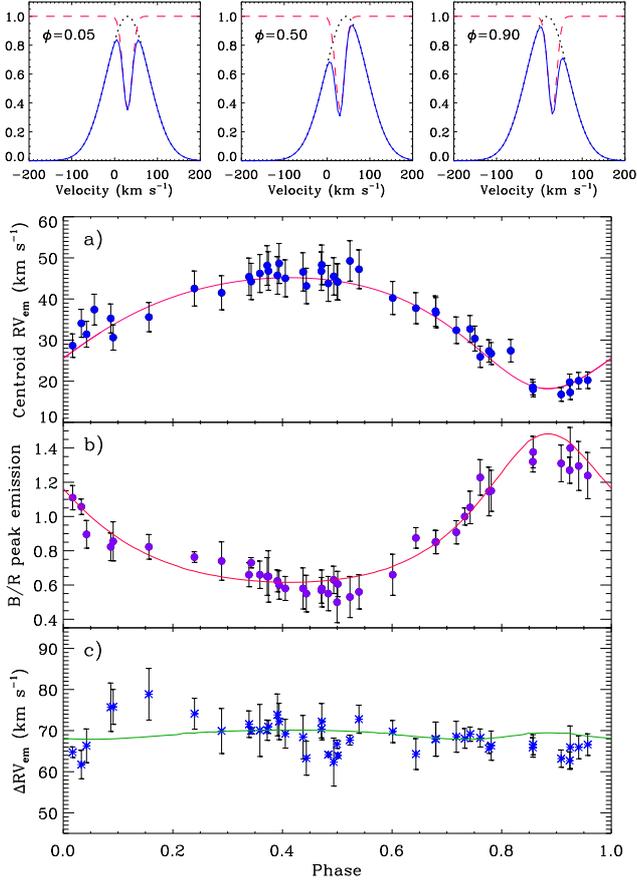}
\vspace{-0.3cm}
\caption{
\textit{Upper three panels:} Synthetic \ion{Fe}{ii} line profiles (full lines), at three orbital phases ($\phi$), resulting from the superposition of a ``moving'' broad Gaussian  emission (dotted lines) and a ``static'' narrower Gaussian absorption.
\textit{Lower panels:} Radial velocity $RV_{\rm em}^{\rm S}$~($a$),   and $B/R$ intensity ratio ($b$) of the \ion{Fe}{ii} emission lines (dots) with the result of the geometrical model superimposed with full lines.
\textit{ panel c :} The velocity separation (asterisks) of the red--blue  peaks. ($RV_{\rm em}^{\rm R}$ -- $RV_{\rm em}^{\rm B}$) with the result of our model superimposed with a full line.      }
\label{fig:model_FeII}
\end{figure}

We also plotted the blue and red peak RVs individually. The
situation is much more complicated here. Sometimes the lines are
only asymmetric with one of the peaks missing. In other cases more
than two peaks are present simultaneously in all the lines making
the measurement much more uncertain. The results are presented in
Table~\ref{tab:rv} (columns from 8 to 11) and in
Fig.~\ref{fig:RV}-c. These RVs show small-amplitude scattered
variations with only a possible modulation with the orbital 60--day
period. However, the velocity difference seems to be constant except
for a few cases around phase 0.15 (see bottom panel of
Fig.~\ref{fig:model_FeII}).

Given the unclear situation for the RV of the double peaks, we
decided to additionally measure the equivalent
width (EW) of the emission lines and the centroid of each line without
fitting its peaks but treating it as a single feature. To do that, we used
the {\sc splot} task in IRAF. The RVs, calculated from the line centroids,
are listed in Table~\ref{tab:rv} ($RV_{\rm em}^{\rm S}$, columns 12 and 13), and
displayed in Fig.~\ref{fig:RV}-c and Fig.~\ref{fig:model_FeII}-a. A
clear, positive correlation with the absorption lines RV curve is now
evident, although with an amplitude smaller than that of $RV_{\rm
abs}$ and RVs ranging from +15 to +50 km\,s$^{-1}$. The
periodogramme analysis applied to the RV of the emission lines
practically gives the same period as for the absorptions. The fit of
the $RV_{\rm em}^{\rm S}$ curve with an eccentric orbit provides the values
reported in Table~\ref{tab:orb}.
This solution is shown as a full line in Fig.~\ref{fig:model_FeII}-a.
The cross-correlation of $RV_{\rm em}^{\rm S}$ and $RV_{\rm abs}$
indicates that the former precedes the latter by a very small (and
likely insignificant) phase shift $\Delta\phi\approx -0.03$
(corresponding to $\approx 10\degr$).

\begin{figure}[htp]
\hspace{-0.4cm}
\includegraphics[width=11cm]{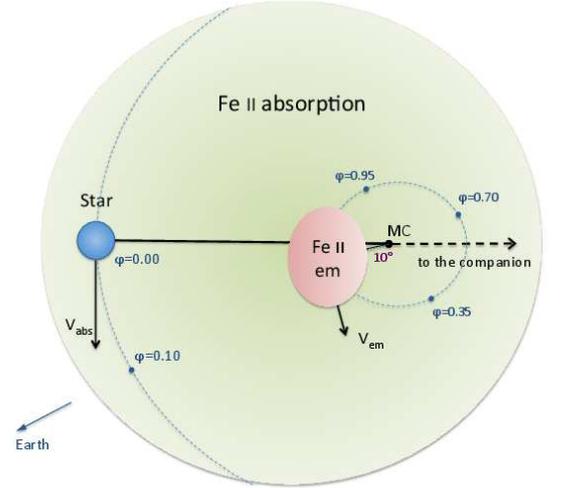}
\vspace{-0.3cm} \caption{Sketch of the system geometry with the emitting and absorbing \ion{Fe}{ii} regions.
The orbit of the star and that of the corotating \ion{Fe}{ii} emitting region are depicted with dotted lines and
the phases of conjunctions and quadratures have been also marked. The size of the star and the
emitting region are not to scale.} \label{fig:sketch}
\end{figure}

The same orbital period and the very similar $\gamma$ velocity
suggest that the bulk of the emission is coming from a region
corotating with the binary system but much closer to the center of
mass (MC) than the source of the absorption lines. A simple geometrical
sketch of the source of the \ion{Fe}{ii} emission is depicted in
Fig.~\ref{fig:sketch}.

With this simple geometrical model we also tried to understand the
orbital modulation of the $B/R$ peak ratio.  We propose a hypothesis
that the double-peaked shape of the permitted \ion{Fe}{ii} emission lines
is a result of an extra-absorption superimposed on a pure emission profile. If the two profiles have
different full width at half maximum (FWHM) and velocities, a
changing blue/red peak asymmetry in phase with the
orbit can be reproduced.
The continuous line overplotted on the observed $B/R$ curve in
Fig.~\ref{fig:model_FeII}-b is derived from the synthetic double-peaked profiles obtained by
the superposition of a broad emission (a  Gaussian with FWHM=115\,km\,s$^{-1}$)
moving according to the  $RV_{\rm em}^{\rm S}$ curve (full line in Fig.~\ref{fig:model_FeII}-a),
and a narrower absorption Gaussian (FWHM=25\,km\,s$^{-1}$) at the MC velocity.

As indicated by the $RV_{\rm em}^{\rm S}$ semi-amplitude of only 13.5\,km\,s$^{-1}$, the emitting
region must be rather compact (or highly centrally condensed) and should be located
much closer to the MC of the system than the star which is the source of the observed
absorption lines. As a simple speculation, if the location of this emitting region is close
to the Lagrangian point L$_1$, one may suggest that it is related to a denser part of
the stellar wind funneled through L$_1$ towards the unseen
companion. Conversely, the excess absorption shows a constant
intensity and velocity, so it must be caused by a quite
large and homogeneous structure, which is always projected
onto the emission produced by the compact region. One could argue
then that the source of the excess absorption is a circumstellar
envelope or shell.

Although this is a  simplified model, it allows us to explain the observed behavior of the permitted
\ion{Fe}{ii} emission lines and could be taken as a starting point
for more sophisticated models.

The EWs of the measured emission lines show no evidence for periodicity.
This supports the hypothesis of a purely ``geometrical'' effect
produced by a static and nearly homogeneous absorbing envelope over the compact
emitting region.

 \subsection{Forbidden emission lines}

Similarly to the permitted emission lines, metallic forbidden lines
have double-peaked profiles that are not always well structured.
Three peaks are clearly present in some spectra. Most of them are
blended with other emissions, so that we could not obtain definite
results. The final list of [\ion{Fe}{ii}] lines to analyze only
includes $\lambda$ \,5627.5, $\lambda$\,5376.5, and
$\lambda$\,5158.8 \AA.

As already described in \cite{Murat08}, forbidden lines are narrower
than permitted lines: a mean FWHM of 55\,$\pm$\,8 km\,s$^{-1}$ is
found from a single Gaussian fitting. The RV curve reproduces the
behavior of the permitted (and therefore of the absorption) lines
though with quite a large scatter, likely due to the small number of
lines and their faintness (see Fig.~\ref{fig:RV}-d). We found
surprising that the RV variation of the [\ion{Fe}{ii}] lines is larger
than that of the permitted lines, indicating a different emission
region. In a very simplified scheme of matter corotating with the
system, this would mean that this region is located farther from the
MC of the system towards the star, which is the source of the
absorption lines. However, this result needs to be confirmed by
further observations with a higher S/N ratio.

\begin{figure}
\includegraphics[width=9.0cm]{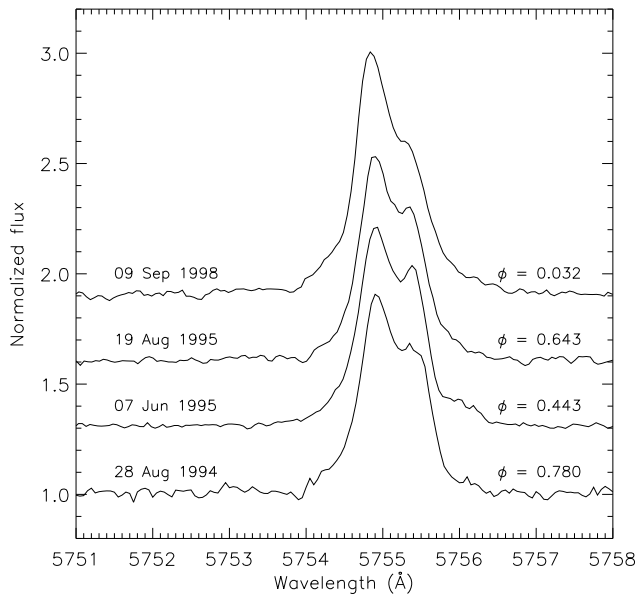}
\caption{The [\ion{N}{ii}]  5754.59\,\AA ~line profiles of the OHP spectra; the observed spectra are displayed
in a heliocentric wavelength scale.
}
\label{fig:niiohp}
\end{figure}

An interesting case is the [\ion{N}{ii}] 5754.59\,\AA\, line which
shows a clear double-peaked profile in all the {\sc Elodie} spectra
($R$=42\,000) with the red peak always fainter than the blue one,
as was also found by \citet{Lobel13}. The mean RV difference between 
the peaks is 27.5 $\pm$ 1.0 km\,s$^{-1}$ (see Fig.~\ref{fig:niiohp}), 
that is much smaller than the peak separation of $\sim$\,70 km\,s$^{-1}$ 
measured for the permitted emission lines (see Fig.~\ref{fig:RV}-c). 
The B/R peak ratio ranges from 1.3 to 1.6. The other spectra have a
lower resolution that does not permit to separate the peaks, however
the {\sc Fresco} spectra indicate an asymmetry of the blue/red side
at the top of the line which is never reversed. Other parameters
characterizing this line remained constant during the long period of
our observations. The results can be summarized as follows: FWHM= 50
$\pm$ 5 km\,s$^{-1}$,  EW=0.8 $\pm$ 0.1 \AA,
 $<$RV$>=28.0\pm$1.3 km\,s$^{-1}$. The latter is very close to the 30\,km\,s$^{-1}$
of the barycentric velocity (see Fig.~\ref{fig:RV}). This could
indicate an emitting region located around the
MC or, most likely, an emitting envelope/shell with a large
extension, encompassing the entire binary system, similar to the one
which we have assumed to give rise to the narrow \ion{Fe}{ii}
absorption.
A very similar behaviour is displayed by the [\ion{N}{ii}] 6583.45\,\AA\  line. 

With the data in our hands, we cannot say whether the extra-absorption is a
feature related to the formation of the [\ion{N}{ii}] lines or it is an effect of the same cloud that
causes the excess absorption in the permitted \ion{Fe}{ii} lines. In
the latter hypothesis, the absence of inversion of the B/R peak
ratio is due to the stability of [\ion{N}{ii}] line RV.

However, this line can mainly form in the large bipolar outflow around MWC\,314
which has been discovered by \citet{Marston08} with narrow-band H$\alpha$ imaging. Since both
the {\sc Elodie} and {\sc Fresco} spectra are taken with optical fibers which encompass a few arcseconds on
the sky around the central star, the double-peaked profiles could be indicative of the approaching and receding
parts of the outflow in the line of sight of the central star. Similar [\ion{N}{ii}] 6583.45\,\AA\  double-peaked profiles with velocity separations
of the same order have been also observed in LBV candidates in the Large Magellanic Cloud by \citet{Weis03}. 

\section{Hydrogen and Helium line profiles }

We also investigated the behavior of the helium and hydrogen lines with the orbital phase.

The H$\alpha$ line with a peak intensity $\sim 20$ in the continuum
units and a full width at zero level of $\sim 70$\,\AA\ is a strong
and broad feature. It is difficult to set the continuum level around
the line and precisely measure its EW. However, its centroid (and
RV) can be defined with a sufficient accuracy. Most of other H and
He lines are either faint, fall near edges of \'echelle orders, or
are blended with other emission lines. We found that the
\ion{He}{i}\,$\lambda$5876 \AA\ and H$\beta$ lines appear in all our
spectra, are strong enough, and unsaturated. Unfortunately, the
other strong helium line in the optical range,
\ion{He}{i}\,$\lambda$6678 \AA, falls in the gap between the last
two \'echelle orders in our {\sc Fresco} spectra. In the same
spectra, H$\gamma$ falls near the edges of two orders, but we could
measure its EW and RV in the  best exposed ones.

The EWs of H$\alpha$, H$\beta$, and H$\gamma$ are in the ranges of
110--140\,\AA, 12--17\,\AA, and 2--5\,\AA, respectively, with no
clear orbital modulation. The same holds true for their RVs which
are scattered between $\approx$\,30 and 45\,km\,s$^{-1}$.
The absence of orbital modulation in both the EWs and RVs of the
Balmer lines suggests that these features are neither related to one
of the stellar components nor are formed in a compact source
corotating with the system, such as the source of \ion{Fe}{ii}
lines. The Balmer lines more likely originate in a large region or
in different regions that occupy a large portion of the circumbinary
space.

\begin{figure}
\includegraphics[width=9.0cm]{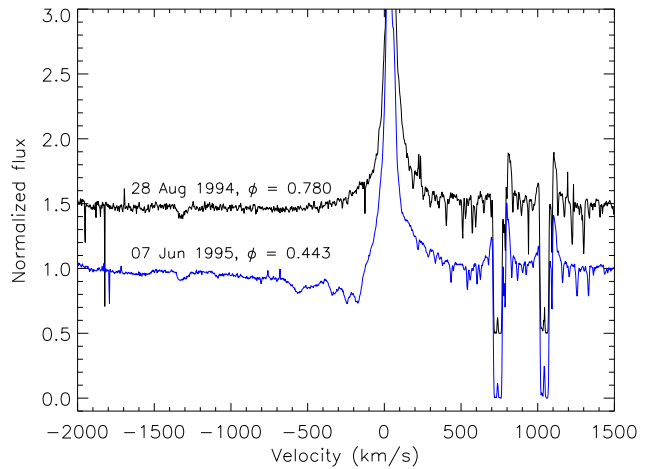}
\caption{OHP-{\sc Elodie} spectra in the \ion{He}{i} region plotted in a scale of heliocentric velocity. The spectra are vertically
shifted for the sake of clarity.}
\label{fig:He_OHP}
\end{figure}

The most interesting behavior shows the profile of the
\ion{He}{i}\,$\lambda$5876 line, which changes from a nearly
symmetric emission to a P-Cygni profile. This is clearly illustrated
by the two OHP-{\sc Elodie} spectra shown in Fig.\,\ref{fig:He_OHP}.
The blue-shifted extra-absorption visible in the spectrum at the
phase $\phi=0.443$ seems to be a result of different components with
velocities from $\sim -$200 \,km\,s$^{-1}$ to $\le
-$1000\,km\,s$^{-1}$.

\begin{figure}
\includegraphics[width=9.0cm]{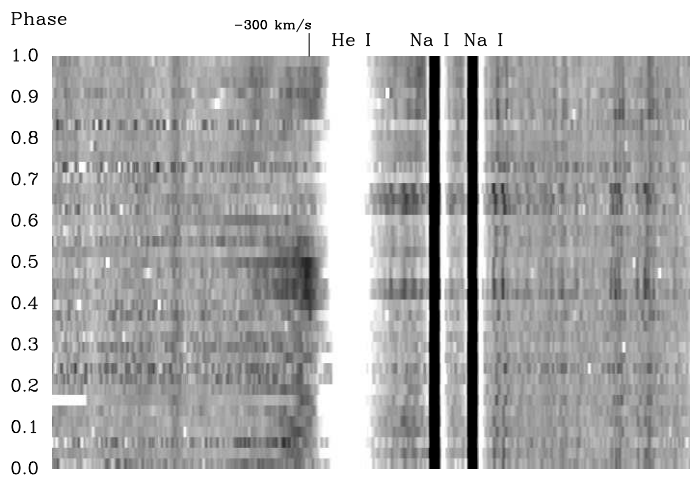}
\caption{Trailed spectrum in the \ion{He}{i}\,$\lambda$5876 region produced with the {\sc Fresco} and Asiago spectra.
The \ion{He}{i} emission peak has been artificially saturated to
emphasize the extra-absorption feature. }
\label{fig:trailed}
\end{figure}

To check if this feature has a random occurrence or it is related to
particular orbital configurations, we have produced a trailed
spectral image with all the {\sc Fresco} and Asiago spectra in the
region around the \ion{He}{i}\,$\lambda$5876 \AA\  line
(Fig.\,\ref{fig:trailed}). It clearly shows that the extra absorption is mainly observed at
phases between 0.40  and 0.55.
Although with a smaller intensity, some extra-absorption is also
observed at phases around $\phi=$\,0.9 and $\phi=$\,0.1 (primary
star in front). This is the configuration where this phenomenon was
observed by \citet{Lobel14}, who ascribed it to wave propagation
linked to the orbital motion near the low-velocity regions of the
stellar wind close to the primary star. Note that, due to the
different ephemeris, the phases in \citet{Lobel14} are shifted by
$\sim -0.15$ compared to ours. They suggest that these absorption
features can also be originated in dynamical wind regions confined
between the two components. Our observations indicate that these
extra absorptions are stronger when the primary star is behind the
``inter-binary'' wind. 
This can be understood if the primary component is filling (or almost filling) 
its Roche lobe, as proposed by \citet{Lobel13} and \citet{Richardson14}, and
transferring mass to the secondary through Roche lobe overflow. 
The strengthening of the extra absorption at these phases
could be due to a gas stream projected over the primary star or to
overdense regions in the stellar wind funneled through L1. Thus, the
source of this phenomenon could be different, both physically and
geometrically, than that originating the \ion{Fe}{ii} emission. This
is in line with the detailed model of asymmetric wind developed by
\citet{Lobel13}. A detailed physical model of the \ion{He}{i}
profile is, however, out of the scope of the present paper.    

%

\section{Concluding remarks}

The large number of high resolution spectra spanning nearly 15 years
has allowed us to confirm the binary nature of MWC\,314 and provide
a very accurate solution for the RV curve obtained from the stellar absorption lines.

The numerous metallic emission lines present in the spectrum very
often display double-peaked profiles with a variable shape.
The RV variations of the permitted \ion{Fe}{ii} emission lines, as derived from their
centroids, are clearly periodic and in phase with the RV curve of
the star that produces the absorption spectrum, although with a much
smaller RV semi-amplitude. The blue/red peak ratio is also strongly
correlated with the orbital period.

We proposed a simple geometrical model, where the source of \ion{Fe}{ii} emission is rather compact
and located near the system MC possibly close to the Lagrangian L1 point. In this model, the central
absorption in the \ion{Fe}{ii} lines forms in a very wide (circumbinary?) region that slowly rotates around
the MC (see Fig.~\ref{fig:sketch}) and is is always projected over the source of emission.
This model allows us to explain the observed variation of both the radial velocity($RV_{\rm em}^{\rm
S}$) and the $B/R$ peak ratio of the permitted \ion{Fe}{ii} lines
fairly well, and it is compatible with a more sophisticated 3D-wind model proposed by \citet{Lobel11}.

The optically-thin [\ion{Fe}{ii}] lines also display double-peaked profiles but their RV variation
has a larger amplitude compared to the \ion{Fe}{ii} ones, indicating that these lines are formed
closer to the primary star and possibly in a circumstellar disk. However since the [\ion{Fe}{ii}] lines are
weaker than those of \ion{Fe}{ii}, this result needs further verification using higher signal-to-noise spectra.

Other emission lines, such as  the [\ion{N}{ii}] 5754\,\AA\  and the Balmer lines
have an almost constant RV which is nearly
coincident  with the $\gamma$ velocity. They likely originate in a
distant, circumbinary region and possibly in the inner parts of the large bipolar outflow discovered by
\citet{Marston08} with narrow-band H$\alpha$ imaging.

The most prominent variable feature observed in the
\ion{He}{i}\,$\lambda$5876 \AA\ line profile is a broad blue-shifted
absorption which is mostly observed at  phases
between 0.40 and 0.55 (primary star behind).
This is likely due to the wind of the primary component, funneled towards
the companion, that is interposed between us and the primary star at
these phases.

In order to continue revealing the nature of the MWC~314 binary system, a new set of high-resolution
and high signal-to-noise ratio spectra, which would densely cover the orbital phases and focus on weak
spectral lines in an attempt to get information about the forbidden lines formation region and search for
the elusive secondary companion, is needed.
Contemporaneous multicolor photometry would help to associate spectral and brightness variations.

\begin{acknowledgements}
The authors thank the staff of the OAC for the technical
support during the observations; IRAF is distributed by the NOAO
which is operated by AURA under contract with NFS.
This project was partially supported by the INAF grant 2007 ``Luminous Blue Variable
phenomenon: towards a better understanding of massive stars
evolution''. We thank the anonymous referee for the useful comments and suggestions.
\end{acknowledgements}

\end{document}